\documentclass[]{article}
\usepackage{emulateapj,onecolfloat,graphicx}

\def\eg{{\it e.g.}}
\def\etal{{\it et al.}}

\def\DF{{\sc DF}}

\def\ba{\;\pmb{\mit a}}
\def\br{\;\pmb{\mit r}}
\def\bv{\;\pmb{\mit v}}
\def\pmb#1{\setbox0=\hbox{$#1$}%
  \kern-0.25em\copy0\kern-\wd0
  \kern.05em\copy0\kern-\wd0
  \kern-0.025em\raise.0433em\box0}

\slugcomment{\it Rutgers Astrophysics Preprint \#488, Submitted to ApJ}

\begin{document}

\twocolumn[
\title{Halo density reduction by baryonic settling?}
\author{J. R. Jardel}
\and
\author{J. A. Sellwood}
\affil{Rutgers University, Department of Physics \& Astronomy, \\
136 Frelinghuysen Road, Piscataway, NJ 08854-8019 \\
{\it jjardel:sellwood@physics.rutgers.edu}}

\begin{abstract}
We test the proposal by El-Zant \etal\ that the dark matter density of
halos could be reduced through dynamical friction acting on heavy
baryonic clumps in the early stages of galaxy formation.  Using
$N$-body simulations, we confirm that the inner halo density cusp is
flattened to $0.2$ of the halo break radius by the settling of a
single clump of mass $\ga 0.5\%$ of the halo mass.  We also find that
an ensemble of 50 clumps each having masses $\ga 0.2\%$ can flatten
the cusp to almost the halo break radius on a time scale of
$\sim9\;$Gyr, for an NFW halo of concentration 15.  We summarize some
of the difficulties that need to be overcome if this mechanism is to
resolve the apparent conflict between the observed inner densities of
galaxy halos and the predictions of LCDM.
\end{abstract}

]

\section{Introduction}
\label{intro}
The LCDM model for structure formation in the universe has been very
successful on large scales (\eg\ Springel, Frenk \& White 2006), but
has run into a number of difficulties on galaxy scales.  One
difficulty is that the predicted dark matter density in the halos of
bright galaxies appears to be greater than that inferred from the best
observational data (\eg\ Weiner \etal\ 2001; Dutton \etal\ 2007) or
from dynamical friction constraints (Debattista \& Sellwood 2000); see
Sellwood (2008b) for a review.

A number of possible solutions have been suggested.  Binney, Gerhard
\& Silk (2001) and others suggest that the halo density can be reduced
by feed-back from star formation, but Gnedin \& Zhao (2002) show that
the density cannot be reduced by more than a factor two, even for the
most extreme feedback, unless the baryons are unreasonably
concentrated.  Weinberg \& Katz (2002) propose that dynamical friction
between a bar and the halo can reduce the central density, but
Sellwood (2008a) shows that significant reductions require extreme
bars and remove a large fraction of the angular momentum from the
baryons.

A third possible solution was proposed by El-Zant, Shlosman \& Hoffman
(2001, hereafter EZ01), who suggested that dynamical friction between
heavy gas clumps and dark matter would transfer energy to the dark
matter, thereby reducing its inner density.  

Essentially EZ01 propose a process of mass segregation that allows the
baryonic matter to displace the dark matter.  If it works, it would
amount to baryon settling with halo {\it de}-compression, as Dutton
\etal\ (2007) argued would be required to improve agreement of LCDM
predictions with scaling relations of $\sim L_*$ galaxies.  Kassin, de
Jong \& Weiner (2006) also remark that rotation curves could be more
easily reconciled with LCDM predictions if halo compression could
somehow be avoided.  Tonini, Lapi \& Salucci (2006) use the idea to
predict rotation curves for galaxies.  Mo \& Mao (2004) apply the
mechanism to pre-process small halos in the very early stages of
structure formation.

A similar, but significantly distinct, idea was tested by Gao \etal\
(2004), who argue that some unspecified attractor mechanism causes
violent relaxation processes to reset the collisionless mass profile
of the combined dark and baryonic matter.  Since violent mergers
disrupt disks, Gao \etal\ clearly invoke a different process from the
slow frictional in-spiral of baryonic clumps proposed by EZ01.

EZ01 suppose that the baryons collect into dense mass clumps which
then sink to the center by dynamical friction.  They assume: (1) the
clumps are small enough they do not collide and (2) are sufficiently
tightly bound that they are not tidally disrupted.  They also
implicitly assume that the dense clumps (3) maintain their coherence
independent of any internal physical processes (such as possible star
formation) and (4) they require the mass clumps to contain little dark
matter.

The in-spiral of massive clumps has been studied extensively and we do
not attempt a complete list of references.  Van den Bosch \etal\
(1999) estimate the infall rate of dark matter sub-clumps through
dynamical friction.  Ma \& Boylan-Kolchin (2004) simulate a mass
spectrum of satellites each composed of particles moving within a main
halo; they find that the combined density profile of the inner halo
and disrupted satellites can either steepen or flatten, depending on
the properties of the clumps.  Arena \& Bertin (2007) study both the
strength of the friction force and the effect on the density profile
for both cuspy and cored initial galaxy profiles.

Here we undertake a direct test of the proposal by EZ01, using
self-consistent $N$-body simulations, in order to determine the clump
mass required to effect a significant density reduction.  We first
derive the settling time-scale for single massive clumps in cuspy
halos, and then present simulations containing an ensemble of heavy
clumps in a background of light halo particles.  We discuss the
physical plausibility of their assumptions in the concluding section.

El-Zant \etal\ (2004, hereafter EZ04) apply the same idea to account
for the flattened density profile reported by Sand \etal\ (2002) in
the galaxy cluster MS 2137-23.  In this context, they identify the
galaxies themselves as the heavy baryonic clumps that settle to the
cluster center.  Both they, and Nipoti \etal\ (2004), present
simulations to show that the in-spiral of galaxies can cause the
background dark matter density profile to flatten, although both
studies neglect the DM halos attached to the infalling galaxies.  In
an appendix, we show that our simulations are in good agreement with
those of EZ04 for the same problem.

\section{Model and method}
In this work we adopt the NFW halo density profile (Navarro, Frenk \&
White 1997):
\begin{equation}
\rho(r) = {\rho_s r_s^3 \over r(r+r_s)^2},
\end{equation}
where $r_s$ and $\rho_s$ respectively set the radius and density
scales.  Since the mass is logarithmically divergent at large radii,
we adopt $M_* = 4\pi \rho_s r_s^3$ as a convenient mass scale; for
reference, $M_*$ is the mass enclosed within $r \simeq 5.305r_s$ and
the mass enclosed within a sphere of $15r_s$ is $\simeq 1.835 M_*$.
An isotropic distribution function (\DF) for the NFW halo can be
obtained by Eddington inversion (Binney \& Tremaine 2008, eq.~4-46).
We limit the radial extent of the halo by eliminating all particles
with energy $E>\Phi(r_{\rm max})$, where $\Phi$ is the gravitational
potential of the untruncated halo and $r_{\rm max}=15r_s$.

In most of our work, we calculate the self-forces between the halo
particles using a surface harmonic expansion on a spherical grid
(Sellwood 2003).  We employ a radial grid with 300 zones, and
typically expand up to $l_{\rm max} = 4$ for the field of the halo
particles.  We recenter the spherical grid on the densest point in the
halo at frequent intervals using the method described by McGlynn
(1984).  As Zaritsky \& White (1988) find that results with this type
of code can depend upon the recentering mechanism, we show in the
appendix that results from our preferred method agree well with those
using a Cartesian grid where recentering is not needed.

We augment the grid-determined acceleration of a light (halo) particle
at position $\br$ by the attraction of $N_h$ heavy, softened point
masses at positions $\br^\prime$:
\begin{equation}
\ddot{\br} = -G \sum_{N_h} M_h {\br - \br^\prime \over |\br -
\br^\prime|^3} g\left({|\br - \br^\prime| \over \epsilon} \right),
\label{accl}
\end{equation}
where $M_h$ is the mass of each heavy particle and $\epsilon$ the
softening length.  The acceleration of each heavy particle is
\begin{equation}
\ddot{\br}^\prime = -G \sum_{N_l} M_l {\br^\prime - \br \over |\br -
\br^\prime|^3} g\left({|\br - \br^\prime| \over \epsilon} \right),
\label{acch}
\end{equation}
where the summation is over all $N_l$ light particles whose masses are
$M_l$, plus the forces from all other heavy particles as given by
eq.~(\ref{accl}).  This strategy is practicable as long as $N_h$ is
not too large.

In this work, we generally adopt a finite softening kernel, such that
\begin{equation}
g(x) = \cases{ x^3(5x^3 - 9x^2 + 5) & $x<1$ \cr 1 & otherwise,\cr}
\label{kernel}
\end{equation}
which is the attraction of a spherical cloud of mass $m$ having
the density profile $\rho(x) = 15m(2x^3 - 3x^2 + 1)/(4\pi\epsilon^3)$,
that declines as a cubic function from a maximum at the center to zero
at $x=1$.  For this kernel the attractive force peaks at $x \simeq
0.564$ and joins smoothly to the inverse square law at $x=1$.  For
comparison with older work only, we have also used traditional Plummer
softening of the form $g^\prime(x) = x^3 / (1+x^2)^{3/2}$, which
weakens forces at all separations.  Values of $\epsilon$ for the cubic
kernel should be larger than for the Plummer kernel, since attractive
forces from equation (\ref{kernel}) are substantially sharper.

Consistent with previous work, we use the term satellite to describe a
single heavy particle orbiting within the halo.  In this case, we set
the satellite in a circular orbit at some initial radius, and share a
net momentum in the opposite direction among the halo particles, such
that the combined linear momentum of the system is zero.

We have verified that our results are insensitive to the time step,
grid size, and halo particle number.  The code conserves total energy
and angular momentum to better than 0.2\% and 0.05\% respectively,
while linear momentum variations never exceed $0.001
(GM^{3}/r_s)^{1/2}$.  As an additional check of our calculations, we
reproduced the result for a ``halo'' represented by an $n=3$ polytrope
from Bontekoe \& van Albada (1987).  The settling time, using 200
times as many particles, agreed with their value to within their
estimated error.

Henceforth, we adopt units such that $G=M_*=r_s=1$.  A numerically
convenient scaling to physical units is to choose $M_* = 7.6 \times
10^{11}\;$M$_\odot$ and $r_s = 15\;$kpc, which sets the unit of time
$(r_s^3/GM_*)^{1/2} = 30\;$Myr.  An NFW halo with these parameters has
concentration $c \simeq 15$ and $V_{200} \simeq 134\;{\rm km s}^{-1}$;
the time unit would be longer in lower concentration halos.

\begin{figure}[t]
\includegraphics[width=\hsize,angle=0]{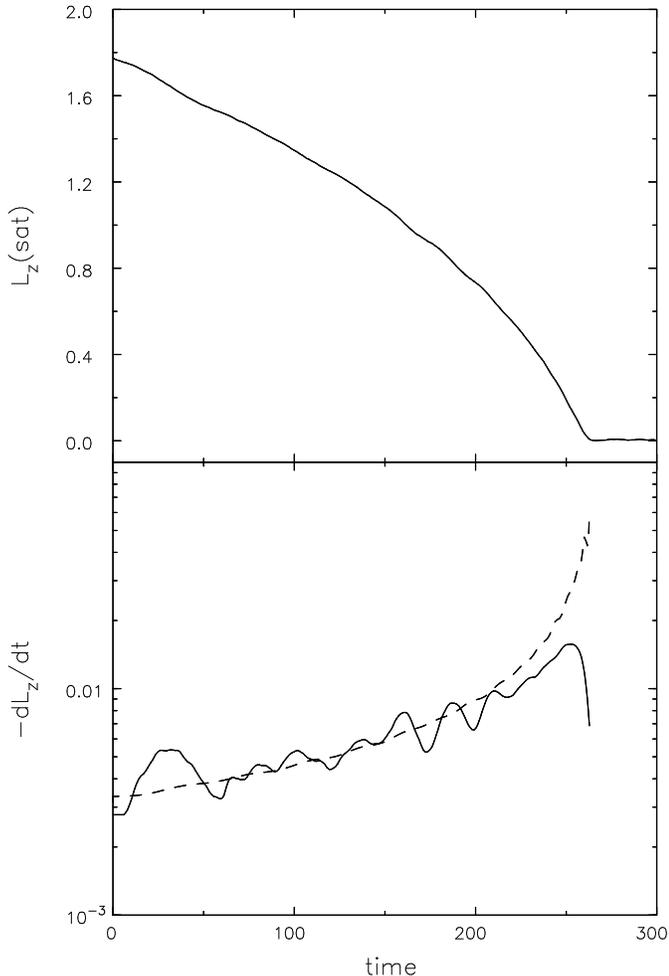}
\caption{\footnotesize Above, the time evolution of the satellite's
orbital angular momentum in the fiducial run.  Below, the frictional
torque measured from the simulation (solid) and predicted by
eq.~\ref{Chandra} (dashed) for $\ln\Lambda=2.5$.}
\label{decay}
\end{figure}

\section{A single satellite}
We begin by revisiting the orbital decay of a single massive
satellite.  The problem of dynamical friction on a heavy particle
orbiting in a spherical system of light particles has been worked on
extensively.  Our purpose here is simply to determine the settling
time of a softened heavy particle in a cuspy halo, before going on to
study the behavior of a collection of such particles in \S\ref{multi}.

Our fiducial run uses a satellite of mass $M_h=0.01M_*$ with
$\epsilon = 0.15r_s$ in an NFW halo, with an initially circular orbit
started at $r= 4 r_s$.  We represent the halo with $10^6$ particles,
use a grid with 300 radial shells, expand up to $l_{\rm max} = 4$ and
employ a basic time step of $0.005 (r_s^3/GM)^{1/2}$.  While the decay
rate of the satellite orbit is independent of most numerical
parameters, it is marginally increased, by $\sim 1$\%, when we
increase $l_{\rm max}$ to 10, and it also depends weakly on the
softening length adopted for the satellite, as discussed below.

\subsection{Settling time}
\label{tscale}
Figure~\ref{decay} (upper) shows the time evolution of the angular
momentum of the satellite.  The rate of loss due to dynamical friction
accelerates as the satellite settles to the center at which point
further change abruptly ceases.  Thus it takes $\sim 260\;$dynamical
times, or 7.8~Gyr, for the satellite to settle to the center of the
NFW halo from its initial radius.  The settling time will be longer in
lower density halos, or if the satellite starts farther out.

\begin{figure}[t]
\includegraphics[width=.9\hsize]{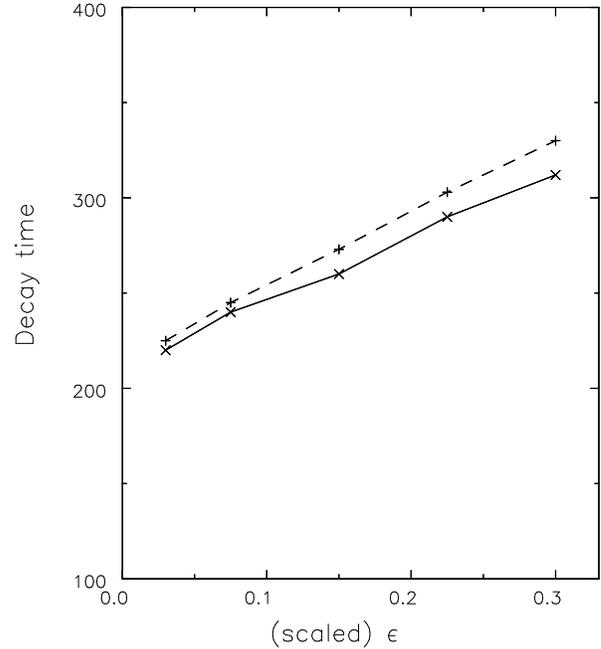}
\caption{\footnotesize The variation of settling time with satellite
softening length.  Results using the cubic softening rule,
eq.~(\ref{kernel}) (points joined by the solid line) are quite similar
to those using a Plummer kernel with $\epsilon/3$ (dashed line).}
\label{dtsoft}
\end{figure}

The well-known formula (Chandrasekhar 1943; Binney \& Tremaine 2008)
for the acceleration $\ba_f$ of a heavy particle moving at velocity
$\bv$ through a sea of light particles of constant density $\rho$ is
\begin{equation} 
\ba_f(v) = -{\bv \over v} 4\pi \ln\Lambda G^2 {M_h \rho \over
\sigma^2} V\left( {v \over \sigma} \right).
\label{Chandra}
\end{equation}
Here $v=|\bv|$, $\ln \Lambda$ is the usual Coulomb logarithm and
\begin{equation}
V(x) =  x^{-2} \left[ {\rm erf}\left({x \over \surd2} \right) -
          \left({2 \over \pi}\right)^{1/2}x e^{-x^2/2} \right];
\end{equation}
the quantity in the square brackets is the fraction of light particles
having speed $<v$, assuming the particles to have a Maxwellian
velocity distribution with dispersion $\sigma$.  While the assumptions
made in the derivation of this formula do not strictly permit it to be
applied to the present problem (\eg\ Tremaine \& Weinberg 1984), it
correctly predicts the scaling with parameters (\eg\ Lin \& Tremaine
1983; Sellwood 2006; Arena \& Bertin 2007).

The lower panel of Fig.~\ref{decay} compares the prediction of formula
(\ref{Chandra}) with our numerical results.  The solid curve shows the
rate of loss of angular momentum by the satellite, while the dashed
curve shows the expected torque, $\br^\prime \times {\ba}_f$, using
local values for the density and velocity dispersion of the halo
particles.\footnote{Measurement of the frictional torque offers a
simpler and more direct comparison with eq.~(\ref{Chandra}) than
through the rate of energy loss used elsewhere (\eg\ Bontekoe \& van
Albada 1987; Arena \& Bertin 2007).}  We treat the Coulomb logarithm,
which is not predicted with any certainty, as a free scaling
parameter; the curve shown is for $\ln \Lambda = 2.5$.  The agreement
is satisfactory until the final plunge of the satellite, where the
empirical torque drops below that predicted.  We have verified that
the frictional torque scales appropriately with the local density and
velocity dispersion, and linearly with the mass of the heavy particle
for $M_h \la 0.05 M_*$.

The low value of $\ln\Lambda$ required to match our data is not purely
a consequence of satellite softening.  The solid line in
Figure~\ref{dtsoft} shows that the time taken for a satellite of mass
$M_h = 0.01$ to settle to the center from a circular orbit at $r=4$
varies approximately linearly with softening length.  As is
reasonable, more concentrated satellites settle more rapidly, but the
variation is by a factor $\sim 1.5$ for a factor ten change in the
softening length, which is a somewhat slower dependence than
$\ln\epsilon$.  Low values of $\ln\Lambda$ are widely reported in
other work (\eg\ Boylan-Kolchin, Ma \& Quataert 2007 and references
therein), and are discussed in the appendix of Milosavljevi\'c \&
Merritt (2001).

\begin{figure}[t]
\includegraphics[width=.99\hsize,angle=0]{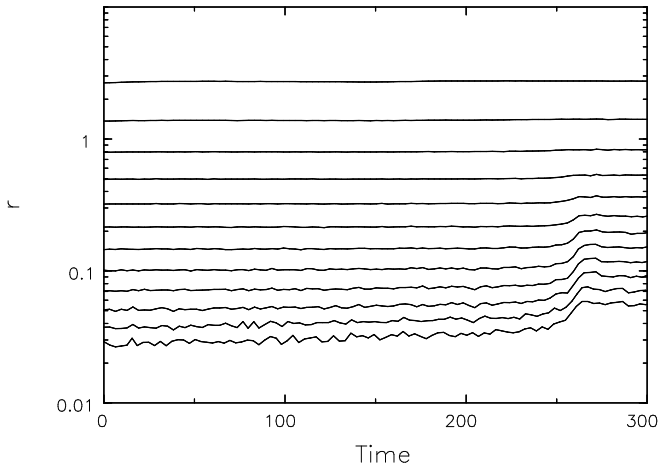}
\caption{\footnotesize The time evolution of the Lagrangian radii in
the fiducial run.  Each curve shows the radius enclosing a fixed
fraction of the halo mass, with the lowest curve corresponding to
$2\times 10^{-4}$, rising by a factor 2 at each subsequent trace.  The
changes are minor until the satellite enters the cusp.}
\label{massfr}
\end{figure}

Fig.~\ref{dtsoft} also compares the results using the cubic softening
kernel (solid line) with those from the more traditional Plummer rule
(dashed line).  The results compare well when the Plummer softening
length is about 1/3 that for the sharper cubic kernel.  For both
softening rules, we find the torque is reasonably well predicted by
eq.~(\ref{Chandra}) for the same $\ln\Lambda$ when $\epsilon$ is
scaled by this factor.

We have run a small number of additional experiments with satellites
on initially non-circular orbits.  The Chandrasekhar friction formula
(eq.~\ref{Chandra}) continues to predict the rate of angular momentum
loss, with the same $\ln\Lambda$.  Generally, the settling time when
the orbit is not strongly eccentric is only slightly longer than that
from a circular orbit with the same initial angular momentum.  More
eccentric orbits take longer to settle, because the satellite spends
more time at radii well beyond its guiding center radius where the
background density is lower.  It should be noted that adding radial
motion at fixed angular momentum, as we discuss here, increases the
satellite's energy, so our conclusion remains consistent with other
work (\eg\ van den Bosch \etal\ 1999).

\begin{figure}[t]
\includegraphics[width=.99\hsize,angle=0]{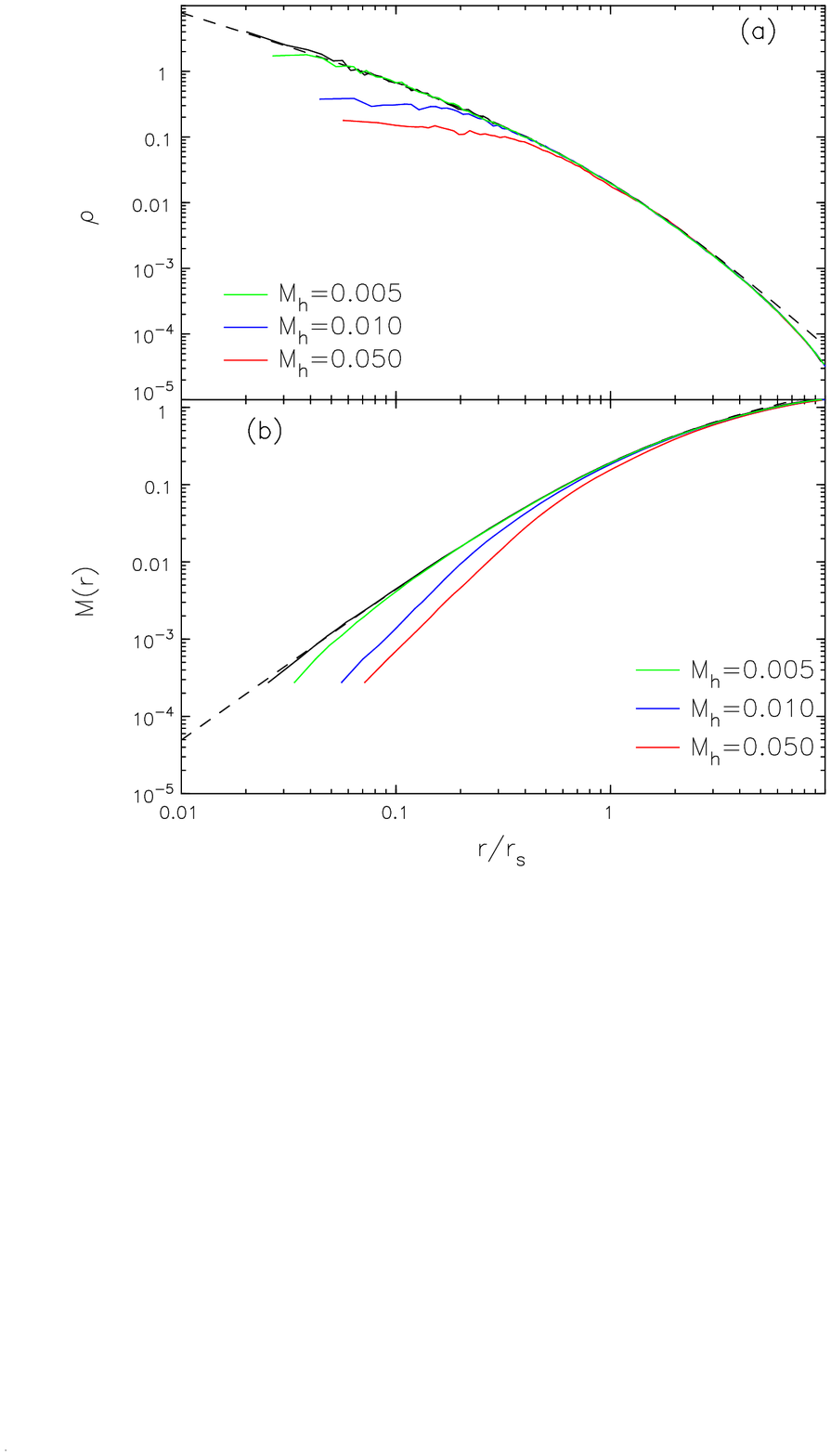}
\caption{\footnotesize The changes in (a) density and (b) mass profile
caused by the settling of a single satellite of 0.5\%, 1\% and 5\% of
the halo mass from an initially circular orbit at $r=4rs$ in an NFW
halo.  The solid lines show the density or mass profiles measured from
the particles at the start (black) and when the satellite has settled
(color).  The dashed lines show the corresponding theoretical NFW
curves.}
\label{massch}
\end{figure}

Our experiments have been confined to spherical dark matter halos,
whereas halos are expected to be mildly triaxial (\eg\ Jing \& Suto
2002).  A mass clump in a triaxial halo may pursue a box orbit (Binney
\& Tremaine 2008, ch.\ 3), causing it to pass through the dense center
from time to time and possibly reducing the settling time (\eg\ Pesce,
Capuzzo-Dolcetta \& Vietri 1992).  However, such orbits are already
eccentric and settle more slowly than quasi-circular orbits.

\subsection{Halo density changes}
As the satellite settles to the center, the halo particles are heated
and the density of the inner halo decreases.  Figure~\ref{massfr},
from our fiducial run, shows the time evolution of the radii
containing many different mass fractions, indicating that the mass
profile undergoes little change until after time 250.  These
Lagrangian radii are always computed from a center at the densest
point.  Since the distance of the satellite from the center falls
below $r_s$ only after $t \sim 250$, most of the change to the mass
profile of the halo occurs during the final plunge of the satellite
through the cusp, in agreement with the conclusions of EZ01.

Figure~\ref{massch} shows the density and mass profiles after the
satellite has settled from the same initial orbit in each case.  The
changes are not large; the cusp is flattened inside $r \la 0.2r_s$ in
our fiducial run (upper panel), where the satellite mass is $0.01M_*$
or $\ga 0.5\%$ of the {\it halo\/} mass.  The changes produced by
heavier and lighter infalling masses differ in the expected sense.
These measurements of the density and mass profiles are, at all times,
from a center located at the densest point.

\begin{figure}[t]
\includegraphics[width=.99\hsize,angle=0]{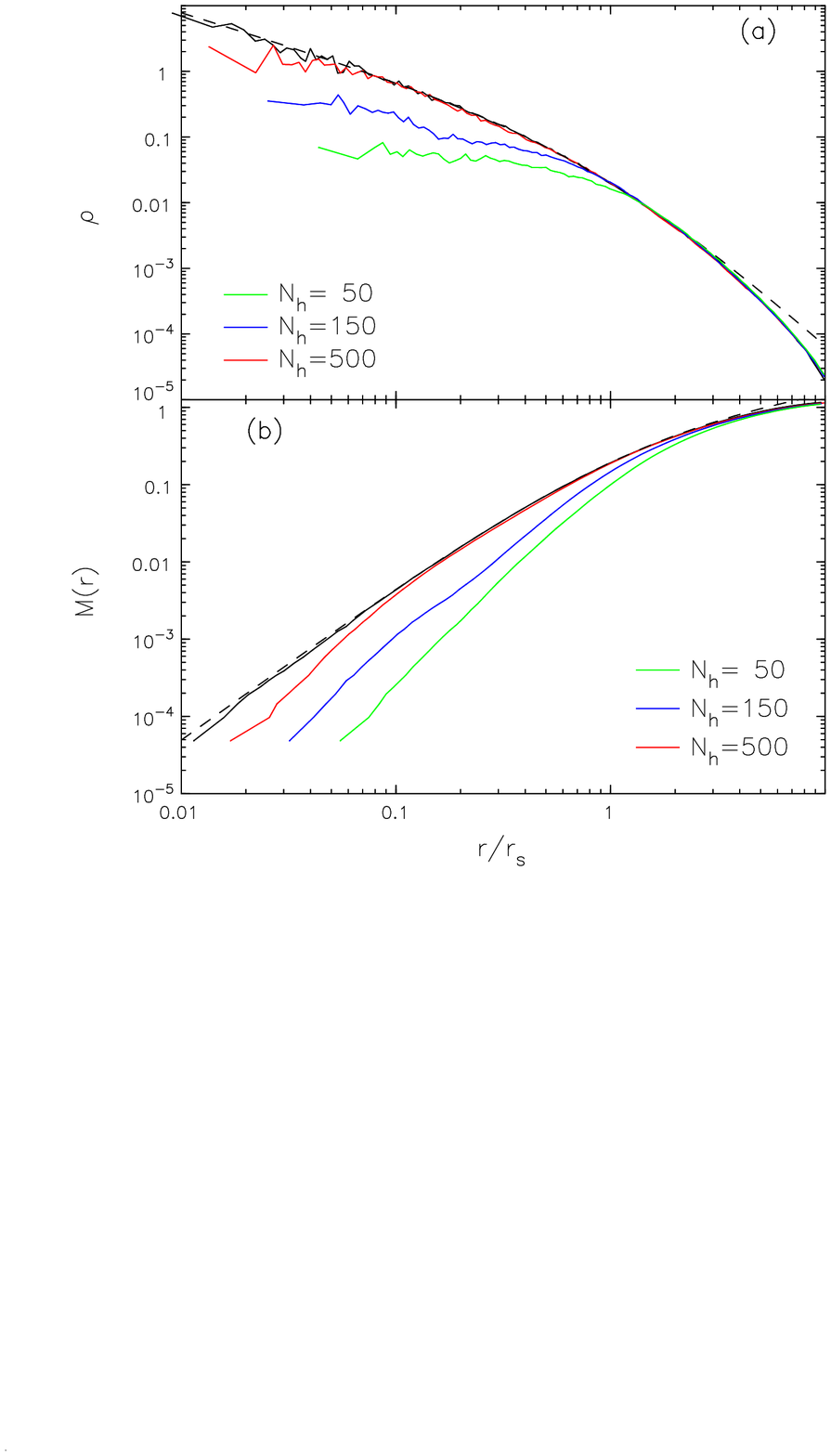}
\caption{\footnotesize The changes in (a) density and (b) mass profile
caused by the settling of $N_h$ heavy particles with total mass of
$0.1M_{200}$, initially distributed at uniform density within a sphere
of radius $4rs$ in an NFW halo.  The solid lines show the density or
mass profiles measured from the particles at the start (black) and
after 9 Gyr (color).  The dashed lines show the corresponding
theoretical NFW curves.}
\label{collect}
\end{figure}

\section{Collection of heavies}
\label{multi}
Dark matter and baryons are well mixed in the early universe.  As
halos form, the relative distributions of the collisionless dark
matter and the collisional baryonic component become more difficult to
predict as shock heating, feedback from star formation, and cold
inflows may all occur (\eg\ Cattaneo \etal\ 2007).  The typical
expectation is that some 5\%--15\% of a total galaxy's mass is
baryonic.

In their models, EZ01 parameterize the clump mass as $M_{200}/\eta$,
where $M_{200}$ is the total (baryonic plus dark) mass of the galaxy.
If the baryon fraction is $f$ and all the baryons are in equal mass
clumps, the number of clumps is therefore $f\eta$.  These authors use
the Chandrasekhar dynamical friction formula to predict the rate at
which energy is removed from the heavy clumps, which they add locally
to the dark matter and adjust its mass profile in response to the
energy added at each radius.

Here instead, we use $N$-body simulations with both heavy and light
particles.  Consistent with their assumptions, we treat the baryonic
mass clumps as heavy softened particles and integrate their motion
through the smooth background halo composed of light, collisionless
particles.

We first simulate their Model 1, which has a baryonic mass of 10\% of
the total enclosed within $15r_s$.  In this model, the baryons are
distributed uniformly inside a sphere of radius $r=4r_s$, while the
dark matter has an NFW distribution.  We therefore redetermine the
initial equilibrium \DF\ for the halo, to take account of the
additional central attraction caused by the baryonic matter.  In our
simulation, the dark matter is represented by $500\,000$ particles
drawn from the new \DF, while we set the isotropic initial velocities
of the 500 equal-mass heavy particles using the procedure described by
Hernquist (1993).

We find some density reduction of the inner NFW profile as the heavy
particles settle.  Figure~\ref{collect} shows the density and mass
profile of the light (dark matter) particles after 300 dynamical
times, or $\sim9\;$Gyr. A small reduction of the dark matter density
is achieved over this time when 500 heavy particles represent the
entire baryonic mass ($f=0.1$).  Naturally, the effect is increased as
the baryonic mass is concentrated into fewer, more massive particles,
since each heavy particle experiences stronger friction and settles
more quickly.

If the baryonic mass clumps are distributed as the dark matter, mass
segregation is even slower.  EZ01 find little evolution with $N_h =
500$ in an NFW halo with the heavies distributed as the NFW density
profile, which we confirm in a simulation with this set-up.  Again we
observe more significant changes when we employ fewer, more massive
heavy particles.

Ma \& Boylan Kolchin (2004) also followed the evolution of a main NFW
halo having a smooth mass distribution, together with $\sim 1000$ mass
clumps having a spectrum of masses.  The total mass of the clumps was
7.02\%, with the most massive being 1.51\%, of the main halo.  They
report that the density of the main halo was substantially reduced
when all the mass clumps were included, but was much less affected in
a separate calculation when the three most massive clumps, which each
had masses $\ga 1\%$ of the main halo, were excluded.  This is in
agreement with our finding (Fig.~\ref{collect}) that the halo density
is reduced more as the masses of the individual heavy particles rise
from 0.02\% to 0.2\% of the main halo.

\section{Discussion and Conclusions}
El-Zant \etal\ (2001) proposed that dynamicl friction on massive gas
clumps could lead to the outward displacement of dark matter. They
estimated the magnitude of the effect on a galaxy halo using the
Chandrasekhar dynamical friction formula.  They followed up (El-Zant
\etal\ 2004) with simulations of a galaxy cluster, in which the heavy
particles were the individual galaxies, finding some density decrease.

We present simulations of softened heavy particles moving under the
influence of gravitational forces within a background NFW halo
composed of a large number of light particles.  We confirm that the
inner halo density is lowered by dynamical friction on heavy
particles, as has been found previously (\eg\ Ma \& Boylan-Kolchin
2004; Arena \& Bertin 2007), and show that larger reductions result
from more massive particles that also settle more rapidly.

We specifically test the model proposed by EZ01, who divided all the
baryons into a number of equal mass clumps that were already somewhat
concentrated towards the halo center.  We find (Fig.~\ref{collect})
that the $\sim 10\%$ baryonic mass fraction must be entirely made up
of $\la 150$ equal clumps if the cusp in the dark matter is to be
flattened to $r \la r_s$ within 300 dynamical times, which scales to
$\sim 9\;$Gyr for a $c=15$ halo.

The principal reason that we observe a more mild density reduction
than predicted by EZ01 is that friction is significantly weaker than
they assumed.  They, and Tonini \etal\ (2006), adopted $\ln\Lambda
\simeq \ln\eta \simeq 8.5$ whereas we find $\ln\Lambda \simeq 2.5$
(\S\ref{tscale}), which causes the mass clumps to settle 2-3 times
more slowly than they assumed, with a corresponding reduction in the
rate energy is given up to the halo.

More massive clumps both spiral in more rapidly and displace more dark
matter (Fig.~\ref{massch}).  Ma \& Boylan-Kolchin (2004) presented a
pair of simulations that revealed a much smaller density change when
the three most massive clumps were eliminated compared with the result
when they were retained.  Thus substantial reductions of the inner
halo density require just a few extremely massive clumps, assuming
they are not disrupted as they settle.

The time scale, in years, varies as $\rho_s^{-1/2}$, and therefore
would be longer in lower concentration halos.  At higher redshifts,
the orbital period is a fixed fraction of the age of the universe at a
constant overdensity (since both scale as the square root of the
density in an Einstein-de Sitter universe).  While halo densities rise
with redshift, their relative over-densities are lower (\eg\ Zhao
\etal\ 2003), and therefore more massive clumps are required to have
any effect in the time available; Mo \& Mao (2004) invoke baryon
clumps having 5\% of the halo mass when $2 \la z \la 5$.

EZ04 argue that the same physical process applies to galaxies in
clusters, and present supporting simulations.  Again adopting their
numerical model and assumptions, we confirm (Appendix) their result
that the density cusp flattens for $r \la 0.2r_s$, where we also
demonstrate that the result is independent of the numerical method.

The physical assumptions underlying these successes are more
questionable, however.  Purely baryonic gas clumps are indeed expected
to form through the Jeans instability as gas cools and settles in the
main halo (\eg\ Maller \& Bullock 2004), but the high-resolution
experiments of Kaufmann \etal\ (2006) show that gas fragments formed
in this way have masses ranging up to $\sim 10^6\;$M$_\odot$ only,
some two orders of magnitude smaller than required to have
interestingly short settling times.

Dark matter sub-halos (\eg\ Diemand, Kuhlen \& Madau 2007) may contain
gas and would spiral in more rapidly.  However, settling of such
sub-halos would bring both baryons and DM into the center, diluting
the desired separation of baryons from dark matter.  The resulting
changes to the overall dark matter density profile will clearly be
less substantial.  When Ma \& Boylan-Kolchin (2004) treat sub-clumps as
collections of particles, they find that the stripped mass is added to
the main DM halo in such a way as approximately to replace the mass
scattered to larger radii by dynamical friction.  (The net effect can
be either a small increase or decrease in the inner dark matter
density, depending on the clump mass spectrum.)  If dynamical friction
is to separate the baryons from dark matter with the desired
efficiency, baryonic mass clumps in sub-halos must somehow be stripped
of their dark matter with high efficiency, without dissolving the gas
clumps themselves.

In the case of the cluster simulation, EZ04 describe the heavy
particles as purely baryonic galaxies.  Here again, the change in the
central dark matter density will not be as substantial when the dark
halos of the galaxies are taken into account, as also noted by Nipoti
\etal\ (2004).

Additional density reduction would be possible if massive baryonic
clumps in the center of the halo could be re-accelerated.  Models of
this kind have been proposed by Mashchenko, Couchman \& Wadsley (2006,
2007), who invoke star formation activity, and by Peirani, Kay \& Silk
(2008), who use AGN jets.  It should be noted, however, that massive
clumps of dense gas are not easily accelerated by these astrophysical
processes (\eg\ MacLow \& Ferrara 1999).  The rapid density reduction
reported by Mashchenko \etal\ (2006) occurs because they employed a
gas clump of $10^8\;{\rm M}_\odot$ within the cusp driven sinusoidally
with a bulk speed that peaked at 18~km~s$^{-1}$.

Most halo density reduction occurs as clumps plunge rapidly within the
cusp (Fig.~\ref{massfr}), as noted by EZ01.  Therefore, clumps that
start out in the cusp fall in more rapidly and are most efficient at
displacing the dark matter.  However, only a small fraction ($\la
5\%$) of the baryons start out in the cusp, if they are distributed as
the dark matter.  The desired halo density reduction would be more
rapid if the baryonic fraction in the cusp could be increased, but it
is hard to see how such an initial mass segregation could have arisen
without compressing the halo.

Thus the challenge, if the proposal by EZ01 is to help solve the
central density problem of dark matter halos, is to find ways to
collect gas into clumps exceeding $\sim 0.5\%$ of the halo mass that
can settle as coherent entities into the center.

\acknowledgments We thank Kristine Spekkens and Frank van den Bosch
for helpful comments on a draft of this paper.  An anonymous referee
and the editor, Chung-Pei Ma, also made constructive suggestions for
improvement.  This work was supported by grants AST-0507323 from the
NSF and NNG05GC29G from NASA.

\vfill\eject

\appendix
\section{Galaxy Cluster Simulations}

El-Zant \etal\ (2004) present results from simulations using a
non-adaptive Cartesian grid covering only the inner part of the dark
matter halo of a cluster of galaxies.  They report that the density of
dark matter declined significantly at radii $<r_s/3$ of the NFW halo
profile after $\sim 10\;$Gyr.  The numerical tests reported in this
appendix reproduce their calculation.

\begin{figure}[t]
\includegraphics[width=\hsize=270]{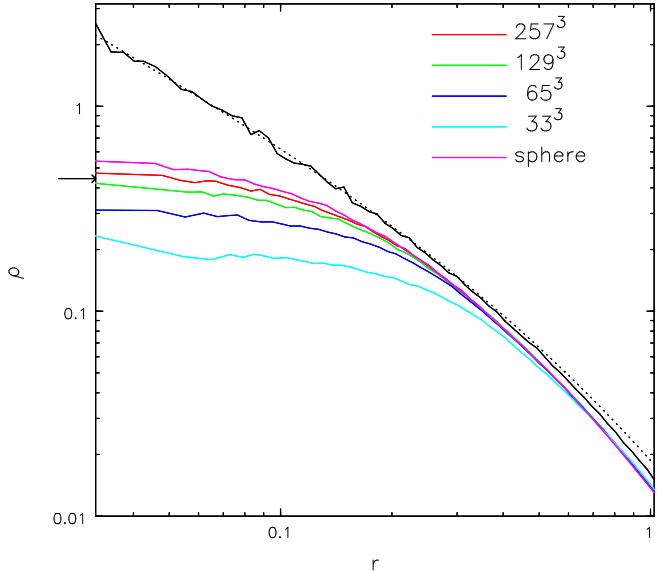}
\caption{\footnotesize The initial (black) and final (color) density
profiles of runs for comparison with EZ04, showing a converge test
with four different Cartesian grids and our spherical grid (magenta).
The core density obtained by EZ04 is marked by the arrow.}
\label{EZcomp}
\end{figure}

Since EZ04 employed a $256^3$ Cartesian grid, they restricted their
calculation to the inner part of the NFW halo.  They set the scale
radius $r_s = 39$ mesh spaces, which implies that particles beyond a
radius $r = 128r_s/39 \simeq 3.28r_s$ could lie outside the grid.
They therefore truncate the NFW halo at $r = 3.33r_s$ and use a
monopole approximation to integrate the motion of particles when they
are off the grid.  Their fiducial simulation uses $9 \times 10^5$
particles to represent the diffuse dark matter, and 90 heavy particles
to represent galaxies, which collectively have a mass that is 6\% of
the total.  Both the dark matter particles and the heavy particles
have initially the same spatial extent and isotropic velocities.

In their calculation, all forces were computed using the grid.  Forces
from a point mass on a Cartesian grid (shown graphically in the
appendix of Sellwood \& Merritt 1994) resemble those from the cubic
spline kernel (eq.~\ref{kernel}) with a softening length of three grid
spaces, or in their case $\epsilon \simeq 3r_s/39$.

Here we reproduce this calculation using our own 3-D Cartesian grid
code (Sellwood \& Merritt 1994), demonstrate numerical convergence,
and compare results from this grid with those from our spherical grid.
As EZ04, we place all particles, both heavy and light, on the $257^3$
grid where we set $r_s = 40$ mesh spaces.  When using other grids we
compute interactions between the heavy particles (galaxies) and the
light particles (dark matter) using the expressions (\ref{accl}) \&
(\ref{acch}), in order to preserve a constant softening length in
physical units ($\epsilon = 0.075r_s$).  We repeated the calculation
with the same physical set-up using grids of size $33^3$, $65^3$, \&
$129^3$ to compute the self-interactions of the light particles, with
the physical length scale $r_s = 5$, 10, \& 20 mesh spaces
respectively.

The results are shown in Figure~\ref{EZcomp}; the initial density
profile is shown by the black line, the profiles after $\sim 10\;$Gyr
on the various grids are shown by the colored lines.  (The colored
lines show time averages over an interval of $\pm0.13\;$Gyr, in order
to obtain smooth curves.)  The magenta line in Fig.~\ref{EZcomp} also
shows the result using the same physical model, but calculated using
our high-resolution spherical grid.  It is clear that the density
reduction diminishes as the Cartesian mesh is refined, although the
final density profile has almost converged at $257^3$ and,
reassuringly, this result agrees quite closely with that from our
spherical grid.  Furthermore, the final central density we obtain is
very similar to that found by EZ04; the inner density in their model
flattens to the value marked by the arrow.

We note that the results shown in Fig.~\ref{EZcomp} can be compared
with $N_h = 150$ curve in Fig.~\ref{collect}.  The masses of the
individual heavy particles are $0.67 \times 10^{-3}$ of the halo in
both cases, but the halo concentrations differ, which affects the
scaling with time.  Since $c=5.45$ for the cluster and $c=15$ for the
galaxies, the square root of the density ratios is $\approx 3.4$.
Therefore the cluster simulation is run for about one third the number
of dynamical times of the galaxy, which also has more heavy particles.
A comparison of the density profiles in the two simulations at equal
numbers of dynamical times reveals very similar density changes, as it
should.

Thus we confirm that a density reduction can indeed be obtained if the
assumptions made by EZ04 hold.  However, they invoke 90 massive
galaxies in the inner part of the cluster, whereas the true number is
probably less (Stanford \etal\ 2002).  Furthermore they implausibly
assume that the galaxies (heavy particles) have no dark halos, as
acknowledged by Nipoti \etal\ (2004); more massive heavy particles
will experience stronger friction, but as some dark matter is likely
to be stripped from each galaxy's halo, the change to the overall dark
matter profile needs to be computed from more realistic simulations
(\eg\ Ma \& Boylan-Kolchin 2004).

\end{document}